\documentclass[prd,preprint,superscriptaddress,nofootinbib]{revtex4}
\usepackage{graphicx}
\usepackage{amsfonts}
\usepackage{latexsym}
\usepackage{amssymb}
\usepackage{amsmath}
\usepackage{epsfig}
\usepackage{color}

\begin{document}

\title{Realization of a spontaneous gauge \\ and supersymmetry breaking vacuum}

\author{Tatsuo Kobayashi}
 \affiliation{Department of Physics, Hokkaido University, Sapporo 060-0810, Japan
}

\author{Yuji Omura}
%\email[]{yujiomur@eken.phys.nagoya-u.ac.jp}
\affiliation{Kobayashi-Maskawa Institute for the Origin of Particles and the Universe, Nagoya University, Nagoya 464-8602, Japan}

\author{Osamu Seto}
 \affiliation{Institute for International Collaboration, Hokkaido University, Sapporo 060-0815, Japan}
\affiliation{Department of Physics, Hokkaido University, Sapporo 060-0810, Japan
}

\author{Kazuki Ueda}
\affiliation{Department of Physics, Hokkaido University, Sapporo 060-0810, Japan
}

%\date{\today}
%
\begin{abstract}
It is one of the major issues to realize a vacuum which breaks supersymmetry (SUSY) and R-symmetry, in a supersymmetric model.
We study the model, where the same sector breaks the gauge symmetry and SUSY.
In general, the SUSY breaking model without gauge symmetry has a flat direction 
at the minimum of F-term scalar potential.
When we introduce U(1) gauge symmetry to such a SUSY breaking model, 
there can appear a runaway direction.
Such a runway direction can be lifted by loop effects, 
and the gauge symmetry breaking and SUSY breaking are realized.
The R-symmetry, that is assigned to break SUSY, is also spontaneously broken at the vacuum.
This scenario can be extended to non-Abelian gauge theories.
We also discuss application to the Pati-Salam model and the SU(5) grand unified theory.
We see that non-vanishing gaugino masses are radiatively generated by the R-symmetry breaking and the gauge messenger contribution.
\end{abstract}

\preprint{EPHOU-17-007}
%\preprint{} 

\vspace*{3cm}
\maketitle

%==================================%
%          Main body               %
%==================================%

\section{Introduction}

It is important to study physics beyond the standard model (SM).
Indeed, several types of extensions have been studied.
One direction of extensions is to assume larger gauge groups, 
e.g. U(1) extension and grand unified theories (GUTs) such as $SU(5)$ and $SO(10)$.
Another direction is supersymmetric extension such as minimal supersymmetric standard model (MSSM).
Supersymmtric gauge-extended models such as supersymmetric GUTs are motivated well by the explanation of the origins of the electroweak (EW) scale and the SM
gauge groups.
In such models, it is an important key how  
gauge symmetries and supersymmetry (SUSY) break down.
It is also interesting to construct models that 
both gauge symmetries and SUSY are broken spontaneously 
by the same sector and their breaking is tightly related with 
each other \cite{Kobayashi:2014iaa,Hirayama:1998rv,Agashe:1998kg,Agashe:2000ay,Bajc:2006pa,Bajc:2008vk,Bajc:2012sm}.

In general, spontaneous SUSY breaking models without gauge symmetries 
have flat direction at the tree-level potential minimum \cite{Ray:2006wk,Sun:2008nh,Komargodski:2009jf} like 
the O'Raifeartaigh model \cite{O'Raifeartaigh:1975pr}.
Such a flat direction could be lifted up by one-loop effects.
SUSY breaking models with $U(1)$ gauge symmetry have been studied as well.
Then, it is found that the $U(1)$ D-term potential does not stabilize the flat direction of the F-term scalar potential.
However, there can appear a runaway direction along which 
the D-term potential becomes vanishing, when D-term is non-vanishing 
at the minimum of F-term scalar potential \cite{Azeyanagi:2011uc}.
Such a runaway direction could be lifted up by one-loop effects 
and a minimum would appear at non-vanishing finite field value.
Then, we could realize both gauge symmetry and SUSY breaking.
Note that R-symmetry, that should be broken to realize finite gaugino masses,
could also be spontaneously broken at the vacuum. 
Thus, we can evade the vanishing gaugino masses that are often predicted in the gauge mediation models \cite{Komargodski:2009jf,Giveon:2009yu,Dudas:2010qg}.

In this paper, we study the above scenario, that is, 
the runaway direction and its lifting to realize both gauge symmetry and SUSY breaking 
by the same sector. At first we discuss the U(1) model, and then extend it to non-Abelian models.
As illustrative models toward realistic GUTs, we discuss the Pati-Salam model \cite{PS} and
the $SU(5)$ GUT \cite{su5,flippedsu5}.
In the Pati-Salam model, the gauge symmetry is $SU(4) \times SU(2)_R \times SU(2)_L$.
The gauge symmetry $SU(4) \times SU(2)_R$ and SUSY are broken at the same time.
That can be also realization of gauge messenger models, which 
can lead a specific spectrum of superpartners \cite{Dermisek:2006qj,Intriligator:2010be,Matos:2010ie}.
Similarly, we see that our SUSY breaking scenario can be applied to the $SU(5)$ GUT
and compare the result with the one in the Pati-Salam model.

This paper is organized as follows.
In section \ref{sec:SUSY-breaking}, we study the SUSY breaking model with $U(1)$ gauge symmetry.
We show that there is a runaway direction and it can be lifted 
by one-loop effects.
In section \ref{sec3}, we extend the U(1) model to non-Abelian models, and 
we apply the above model to the Pati-Salam model and the flipped $SU(5)$ GUT.
Section \ref{conclusion} is devoted to conclusion.

\section{SUSY breaking model}
\label{sec:SUSY-breaking}

In this section, we study flat directions and runaway directions 
in SUSY breaking models with and without $U(1)$ gauge symmetry.
We show that such a runaway direction can be lifted by one-loop effects.
Most of the content in this section is review except lifting the runaway direction 
by one-loop effects.

\subsection{SUSY breaking models without gauge symmetry}

In this section, we review that a generic SUSY breaking model has 
a flat direction at the potential minimum \cite{Ray:2006wk,Sun:2008nh,Komargodski:2009jf} 
like the O'Raifeartaigh model.

We consider renormalizable superpotential $W(\phi_i)$ with $i=1,\cdots, n$.
Here, we use the notation that the chiral superfield $\phi_i$ and its lowest 
component are written by the same letter.
Then, the F-term scalar potential $V_F$ is obtained by  
\begin{equation}
V_F=\sum_i\overline {W_{\bar i}} W_i,
\end{equation}
assuming canonical K\"ahler potential.
Here $W_i$ denotes the first derivative of $W(\phi_i)$ by $\phi_i$, 
and we use a similar notation for higher derivatives.
We assume that the potential minimum is obtained at 
$\phi_i = \phi_i^{(0)}$  and SUSY is broken there.
That is, the stationary condition is satisfied as 
\begin{equation}
\label{eq:VF-minimum}
\frac{\partial V_F}{\partial \bar \phi_j} = \sum_i\overline {W_{\bar ij}}(\phi^{(0)}) W_i(\phi^{(0)}) = 0.
\end{equation}
Since we assume that SUSY is broken, some of $W_i(\phi^{(0)})$ must be non-vanishing.
Actually, the fermion field along the direction $v_i =W_i(\phi^{(0)})$ is massless, 
and corresponds to the Nambu-Goldstone fermion caused by the SUSY breaking.

%%%%%%%%%%%%%%%%%%%%%%%%%%%%%%%%%%%

The mass squared matrix of scalar fields is written by 
\begin{equation}
M^2_B=\left(  
\begin{array}{cc}
\sum_k\overline {W_{\bar ik}}(\phi^{(0)}) W_{kj}(\phi^{(0)}) & 
\sum_k\overline {W_{\bar ijk}}(\phi^{(0)}) W_{k}(\phi^{(0)}) \\
\sum_k\overline {W_{\bar k}}(\phi^{(0)}) W_{ijk}(\phi^{(0)}) &
\sum_k\overline {W_{\bar jk}}(\phi^{(0)}) W_{ki}(\phi^{(0)})
\end{array}
\right).
\end{equation}
Let us evaluate the mass squared along the direction $v_i$, which is 
the superpartner direction of the Nambu-Goldstone fermion.
Its supersymmetric mass is vanishing because of the stationary 
condition (\ref{eq:VF-minimum}).
Thus, the mass squared along this direction is obtained as 
\begin{equation}
\label{eq:mass-2}
\sum_{i,j,k} v_i W_{ijk}(\phi^{(0)}) \overline {W_{\bar k}} v_j + c.c..
\end{equation}
If this mass squared is non-vanishing, 
it can be negative and the vacuum is not stable.
For the vacuum to be stable, the above value should vanish, i.e.,
\begin{equation}
\sum_{j,k}  W_{ijk}(\phi^{(0)})  \overline {W_{\bar k}}(\phi^{(0)}) \overline {W_{\bar k}}(\phi^{(0)}) = 0.
\end{equation}

Now, let us consider the following direction, $z$:
\begin{equation}
\label{eq:flat-direction}
\phi_i = \phi_i^{(0)} + z \, \overline {W_{\bar i}}(\phi^{(0)}).
\end{equation}
By use of the above results, we find
\begin{equation}
W_i(\phi^{(0)}_i + z \overline {W_{\bar i}}(\phi^{(0)}) )=W_i(\phi^{(0)}_i).
\end{equation}
That is, the F-term scalar potential is flat,
$V_F(\phi^{(0)}_i) = V_F(\phi^{(0)}_i + z \overline {W_{\bar i}}(\phi^{(0)}) )$,
along the above direction in Eq. (\ref{eq:flat-direction}).

Such a flat direction could be lifted by the one-loop effects \cite{Coleman:1973jx}
\begin{equation}
V_{1-loop}(X) = \sum(-1)^F \frac{1}{64 \pi^2} {\cal M}^4_i \ln({\cal M}^2_i/\Lambda^2).
\end{equation}
In the regime that the magnitude of soft SUSY breaking term is much smaller than 
the field value, the full potential including loop effects could be written by \cite{Intriligator:2008fe}
\begin{equation}
V_{eff} \approx \sum |W_i|^2Z_i^{-1},
\end{equation}
where $Z_i$ denotes the wave-function renormalization of $\phi_i$.

\subsection{SUSY breaking model with U(1) gauge symmetry}
\label{sec;SUSYbreaking}

Here, we study a SUSY breaking model with U(1) gauge symmetry.
We show that the flat direction in the previous section is still flat even including 
the D-term potential, but there can appear a runaway direction, 
along which certain fields go to infinity, i.e., $\phi \rightarrow \infty$, 
and  the potential becomes lower \cite{Matos:2009xv,Azeyanagi:2011uc}
(see also Ref. \cite{Azeyanagi:2012pc}).

The full scalar potential is given by 
\begin{equation}
V = V_F + V_D,
\end{equation}
with 
\begin{equation}
V_D = \frac{g^2}{2}D^2, \qquad D = \sum_i q_i |\phi_i|^2,
\end{equation}
where $q_i$ denotes the U(1) charge of $\phi_i$.

First, we show useful relations among the superpotential, D-term and 
their derivatives. 
The superpotential must be  invariant under the U(1) gauge transformation, 
$\phi_i \rightarrow \phi_i + i \varepsilon q_i \phi_i$.
This leads the relation,
\begin{equation}
\label{eq:F-D}
\sum_i W_i q_i \phi_i = \sum W_i D_{\bar i} =0.
\end{equation}
Its derivative by $\phi_j$ is written as
\begin{equation}
\sum_i W_{ij} q_i \phi_i + W_j q_j = \sum (W_{ij} D_{\bar i} + W_iD_{\bar i j}) =0.
\end{equation}
We also obtain 
\begin{equation}
\label{eq:F-D-2}
\sum_j q_j|W_j|^2+\sum_{i,j} \overline W_jW_{ij} D_{\bar i} =0.
\end{equation}

In addition, the stationary condition is written as 
\begin{equation}
\frac{\partial V}{\partial \phi_j}=\frac{\partial (V_F+V_D)}{\partial \phi_j}=
\sum \overline W_i(\phi^{(0)}_i) W_{ij}(\phi^{(0)}_i) + g^2 D(\phi^{(0)}_i) D_j(\phi^{(0)}_i) =0.
\end{equation}
By use of this, we can obtain the following relation:
\begin{equation}
\sum_j q_j |W_j(\phi^{(0)}_i)|^2 - g^2 D \sum_j |D_j(\phi^{(0)}_i)|^2 =0 .
\end{equation}
This relation implies that the D-term is non-vanishing only if 
at least one charged field has non-vanishing F-term.
Otherwise, the D-term vanishes.
When $D$ is vanishing, the structure of the full potential is the same as one of $V_F$.

The mass squared matrix of the scalar fields $\phi_i$ is written by
\begin{equation}
M^2_B=\left(  
\begin{array}{cc}
\sum_k\overline {W_{\bar ik}} W_{kj} +g^2 (D_{\bar i}D_j+D D_{\bar i j})& 
\sum_k\overline {W_{\bar ijk}} W_{k} + g^2 D_{\bar i}D_{ \bar j} \\
\sum_k\overline {W_{\bar k}} W_{ijk} + g^2 D_i D_j &
\sum_k\overline {W_{\bar jk}}W_{ki} +g^2 (D_{ i}D_{\bar j}+D D_{i \bar j})
\end{array}
\right).
\end{equation}
It is found that when both $D=0$ and $W_i \neq 0$ are satisfied,  
the direction $v_i$ has the same mass squared as Eq. (\ref{eq:mass-2}).
Using Eq. (\ref{eq:F-D}), we can show that the direction $z$ in Eq. (\ref{eq:flat-direction}) 
is flat when the minimum of $V_F$, $\phi^{(0)}_i$, satisfies $D(\phi^{(0)}_i)=0$.

Now, let us study the other case that the minimum of $V_F$, 
$\phi^{(0)}_i$, does not satisfy $D(\phi^{(0)}_i)=0$.
Obviously, we find that $V (\phi^{(0)}) $ is larger than $V_F (\phi^{(0)})$ because of the non-vanishing D-term.
We examine the value of $D$ along the direction in Eq. (\ref{eq:flat-direction}):
\begin{eqnarray}
D &=&  \sum_i  q_i |\phi^{(0)}_i + z \overline W_i(\phi^{(0)})|^2 \nonumber \\ 
&=& \sum_i q_i |\phi^{(0)}_i |^2 +|z|^2\sum q_i|W_i (\phi^{(0)})|^2 + z\sum_i q_i \phi^{(0)}_i W_i(\phi^{(0)}) + c.c..
\end{eqnarray}
The third term vanishes because of the relation (\ref{eq:F-D}).
We also find that $\sum q_i|W_i (\phi^{(0)})|^2$ is vanishing according to the relations (\ref{eq:F-D-2}) and (\ref{eq:VF-minimum}).
This result implies that the potential, $V_F +V_D$, is flat along 
the direction defined in Eq. (\ref{eq:flat-direction}) even adding non-vanishing D-term.

On the other hand, we can show that there is the following runaway direction:
\begin{equation}
\phi_i = \phi_i^{(0)} + z_\infty \overline W_i(\phi^{(0)}) +  \frac{c_i}{\bar z_\infty} + {\cal O}(z_\infty^{-2}),
\end{equation}
where $c_i$ is constant.
The D-term along this direction is evaluated as 
\begin{eqnarray}
D &=&  \sum_i  q_i |\phi^{(0)}_i + z_\infty \overline W_i(\phi^{(0)}) +  \frac{c_i}{\bar z_\infty}|^2  \nonumber \\ 
&=& \sum_i q_i |\phi^{(0)}_i|^2  +q_i W_i(\phi^{(0)}) c_i  + q_i\frac{\bar c_i\phi_i^{(0)}}{z_\infty} + c.c. 
+ {\cal O }(z_\infty^{-2}) .
\end{eqnarray}
We can choose $c_i$ such that they satisfy 
\begin{equation}
\label{eq:runaway-cond}
 \sum_i q_i |\phi^{(0)}_i|^2  +q_i W_i(\phi^{(0)}) c_i +  c.c. = 0 .
\end{equation}
Then, we find that $D=0$ in the limit $z_\infty \rightarrow \infty$, that is, 
$V \rightarrow V_F(\phi^{(0)})$.
Hence, there is, in general, a runaway direction in the model, 
where common fields contribute to both U(1) gauge symmetry and SUSY breaking.

We have shown that there is a runaway direction in generic model when 
the minimum of the F-term scalar potential corresponds to the non-vanishing D-term.
Such a runaway direction would be lifted by loop effects.
In the next section, we discuss lifting of the runaway direction by using 
an explicit model.

\subsection{A concrete model}
\label{sec;model}

In this section, we study a concrete model that causes SUSY breaking and predicts a runaway direction at the tree level \cite{Azeyanagi:2011uc}.
Our model includes five chiral superfields, $X_0$, $X_\pm$, and $\phi_\pm$.
The superfields $X_+$ and $\phi_+$ ($X_-$ and $\phi_-$) have U(1) charge, $+1$ ($-1$), 
while $X_0$ is neutral.
We write the superpotential,
\begin{equation}\label{eq:W-model}
W = X_0 (f + \lambda \phi_+ \phi_-) + m_1X_-\phi_+ + m_2X_+ \phi_- .
\end{equation}
We also assign R-symmetry to cause SUSY breaking.
The fields, $X_0$ and $X_\pm$, have R-charge 2, while 
$\phi_\pm$ have vanishing R-charge.

The constants, $f$, $\lambda$ and $m_{1,2}$, can be defined as positive real values.
Assuming that $m_{1}m_2 < \lambda f$ is satisfied, the minimum of the F-term scalar potential is given by 
  \begin{equation}
\label{eq:pot-min}
\phi^{(0)}_+ = -\frac{F}{m_1}, \qquad \phi^{(0)}_- = \frac{F}{m_2}, \qquad X^{(0)}_0 = X^{(0)}_+=X^{(0)}_- = 0,
\end{equation}
where $F$ and $F_0$ are defined as
\begin{equation}
F = F_0\sqrt{f\lambda/(m_1m_2) -1}, \qquad F_0 = \frac{m_1m_2}{\lambda}.
\end{equation}
At this minimum, the F-terms are obtained as 
\begin{equation}
\label{eq:WX}
W_{\phi_+}=W_{\phi_-}=0, \qquad W_{X_0} = F_0, \qquad W_{X_+}=-W_{X_-}=F,
\end{equation}
and the F-term scalar potential is written by 
\begin{equation}
V_F(\phi^{(0)})= F_0^2 + 2F^2 = 2\frac{m_1m_2f}{\lambda} - \frac{m_1^2m_2^2}{\lambda^2}.
\end{equation}
Furthermore, the minimum of the F-term scalar potential has the 
following flat direction:
\begin{equation}
\label{eq:flat-ex}
X= zF_0, \qquad X_+ = zF, \qquad X_- = -zF.
\end{equation}
At the minimum of the F-term scalar potential,
the D-term is evaluated as
\begin{equation}
D^{(0)}= \frac{F^2(m_2^2-m_1^2)}{m_1^2m_2^2}, \qquad V_D = \frac{g^2}{2}(D^{(0)})^2.
\end{equation}
Unless $m_1=m_2$, the D-term $D^{(0)}$ does not vanish.
We can confirm that the value of D-term does not change along the direction of Eq. (\ref{eq:flat-ex}).

Based on the discussion in Sec. \ref{sec;SUSYbreaking}, there is a runaway direction in this kind of model.
We investigate the following direction:
 \begin{equation}
\label{eq:flat-ex-2}
X_0= z F_0+ \frac{c_0}{\bar z}, \qquad X_+ = zF+ \frac{c_+}{\bar z}, \qquad X_- = -zF+\frac{c_-}{\bar z}.
\end{equation}
We choose $c_{\pm}$ such that they satisfy 
\begin{equation}
\frac{F^2(m_2^2-m_1^2)}{m_1^2m_2^2} + 2Re(c_++c_-)F =0,
\end{equation}
that corresponds to the condition (\ref{eq:runaway-cond}).
For large $z$, the D-term and the D-term potential behave as
\begin{equation}
D= {\cal O}(z^{-2}), \qquad 
V_D = {\cal O}(z^{-4}).
\end{equation}

On the other hand, the F-terms of $\phi_{\pm}$ behave as
\begin{eqnarray}
 W_{\phi_+} &=& \frac{m_1}{\overline{z}}\left( c_0 \sqrt{f\lambda/(m_1m_2) -1}  +c_-   \right),  \nonumber \\
 W_{\phi_-} &=& \frac{m_2}{\overline{z}}\left( -c_0 \sqrt{f\lambda/(m_1m_2) -1}  +c_+   \right).
\end{eqnarray}
Then, the full scalar potential becomes
\begin{equation}
V =V_F(\phi^{(0)}) + \frac{C}{|z|^2} +{\cal O}(z^{-3}),
\end{equation}
where $C$ is given by
\begin{equation}
C = {m_1^2}\left| c_0 \sqrt{f\lambda/(m_1m_2) -1}  +c_-   \right|^2 
+ {m_2^2}  \left| -c_0 \sqrt{f\lambda/(m_1m_2) -1}  +c_+   \right|^2.
\end{equation}
Thus, this potential has the runaway direction $z \rightarrow \infty$.
The minimum of $C$ is obtained as 
\begin{equation}
C_{min}=\frac{F^2(m_1^2-m_2^2)^2}{4m_1^2m_2^2 (m_1^2+m_2^2)}.
\end{equation}

Now, let us evaluate loop-effects, assuming $|\lambda|^2 \gg g^2$.
We expand $\phi_{\pm}$ around the minimum,
\begin{equation}
\phi_+ = \phi_+^{(0)} + \delta \phi_+, \qquad \phi_- = \phi_-^{(0)} + \delta \phi_- .
\end{equation}
The mass term of $\delta \phi_{\pm}$ in the superpotential is written by 
\begin{equation}
W = X_0 (\frac{m_1 m_2}{\lambda} + \lambda \delta \phi_+ \delta \phi_-)  + \cdots .
\end{equation}
That is, the non-vanishing $X_0$ generates the supersymmetric mass of $\delta \phi_{\pm}$.
In addition, we have the following term in the scalar potential,
\begin{equation}
V = \left |\frac{m_1 m_2}{\lambda} + \lambda \delta \phi_+ \delta \phi_- \right|^2+ \cdots,
\end{equation}
that makes the mass splitting between scalars and fermions of $\delta \phi_{\pm}$.
Then we obtain the one-loop potential,
\begin{equation}\label{eq:V-1loop}
V_{1-loop} = \frac{m^2_1m_2^2}{32 \pi^2} \ln \left ( |X_0|^2/\Lambda^2  \right ) + \cdots =   \frac{m^2_1m_2^2}{32 \pi^2} \ln\left( |z|^2 F_0^2/\Lambda^2 \right )+\cdots .
\end{equation}
Note that the full potential can be written approximately \cite{Intriligator:2008fe,Vaknin:2014fxa}
\begin{equation}\label{eq:V-full}
V = |F_{X_0}|^2 Z_{X_0}^{-1} + \cdots,
\end{equation}
with 
\begin{equation}
Z_{X_0}^{-1} \approx 1 + 2 \gamma_{X_0} \ln |X_0|/\Lambda,
\end{equation}
where $\gamma_{X_0}$ is the anomalous dimension of $X_0$.

At any rate, the above one-loop correction can lift up the runaway direction.
The potential for $z$ can be approximated as 
\begin{equation}\label{eq:V-full-2}
V = \frac{C_{min}}{|z|^2} + \frac{m^2_1m_2^2}{32 \pi^2} \ln \left( |z|^2 F_0^2/\Lambda^2 \right ) +\cdots .
\end{equation}
Then, the stationary condition, $\frac{\partial V}{\partial |z|} =0$, is satisfied at
\begin{equation}
|z^{(0)}|^2 = \frac{32 \pi^2 C_{min}}{m_1^2m_2^2}.
\end{equation}
Thus, by including the one-loop effects, we can obtain the potential minimum with finite vacuum expectation 
values (VEVs), where both U(1) gauge symmetry and SUSY break down.
$X_0$ and $X_{\pm}$ carry the non-vanishing R-charges, so that the R symmetry is also broken at this vacuum.

For simple illustrating estimation, we take the parameters such that $m_1 \gg m_2$.
Then, we can approximately evaluate the VEVs: 
\begin{equation}\label{eq;app1}
|\phi^{(0)}_+| \approx  \frac{m_2}{\lambda}, \qquad |\phi^{(0)}_-|\approx  \frac{m_1}{\lambda},
\end{equation}
and
\begin{equation}\label{eq;app2}
|X_{\pm}^{(0)}| \approx  z^{(0)}F \approx 2 \sqrt 2 \pi \frac{f}{\lambda m_2} .
\end{equation}
Also, the F-terms are approximated by 
\begin{equation}
|W_{X_+}^{(0)}|  = m_1 \phi_-^{(0)} \approx \frac{m_1^2}{\lambda}, 
 \qquad |W_{X_-}^{(0)}| = m_2 \phi_+^{(0)} \approx   \frac{m_2^2}{\lambda}.
\end{equation}
Note that $|X_{\pm}^{(0)}|$ is much larger than $|\phi^{(0)}_+|$ at the obtained SUSY breaking vacuum.
%This condition can be realized by the assumption, $F \gg m^2_{1,2}$.
In this setup, $|W_{X_\pm}^{(0)}| \gg |W_{\phi_\pm}^{(0)}|$ is also predicted. 

Substituting sample values, let us evaluate the parameters quantitatively.
For instance, fixing the parameters at $(F,  \, m_2, \, g)=(10 \times m^2_1,\, 0.5 \times m_1, \, 0.1 )$,
we estimate the SUSY and gauge symmetry breaking scales as
\begin{eqnarray}
&&|X^{(0)}_{\pm}| \approx ( 2.4 \times 10^3) \, m_1, \, |\phi^{(0)}_{+}| \approx 10 \, m_1, \, |\phi^{(0)}_{-}| \approx 20 \, m_1,\nonumber \\
&& |W_{\phi_{+}}^{(0)}|^2+ |W_{\phi_{-}}^{(0)}|^2 \approx  1.6 \times 10^{-5} \, F^2.
\label{eq;evaluation}
\end{eqnarray}
Note that $|z^{(0)}|$ is approximately evaluated as $240/m_1 $ at this reference point, so that $|X^{(0)}_{\pm}|$ becomes large and $|W_{\phi_{\pm}}^{(0)}|$ are suppressed. $F$ denotes the F-terms of $X_{\pm}$ and 
$F=10 \times m^2_1$ corresponds to $\lambda \approx 0.05$.
If we assume that $F$ is much larger than $m^2_{1,2}$, $X^{(0)}_{\pm}$ becomes larger
while $W_{\phi_{\pm}}^{(0)}$ becomes smaller.

It is important to investigate the masses of the fields in the SUSY breaking sectors.
At this reference point, the scalar masses squared normalized by $m^2_1$ are quantitatively estimated as
\begin{equation}
(1.1 \times 10^{-4}, \, 1.4 \times 10^4, \, 1.4 \times 10^4, \, 2.3 \times 10^5,\, 0.7, \, 1.4 \times 10^4 , \,1.4 \times 10^4  ) \times m^2_1.
\end{equation}
In addition, there is a massive mode from the real part of $z$, whose mass is given by the one loop correction in Eq. (\ref{eq:V-full-2}). The imaginary part of $z$ corresponds to the Goldstone boson of the R symmetry.

Note that the superpotential in Eq. (\ref{eq:W-model}) leads only SUSY breaking vacua.
Adding the D-term, we also find a SUSY breaking vacuum with vanishing $X_0$ and $X_{\pm}$
at the tree level. At this vacuum, $\phi_{\pm}$ and the F-terms of $X_{\pm}$ develop the VEVs,
and the SUSY and the gauge symmetry are broken. This vacuum, however, suffers from tachyonic
masses of the sfermions, as discussed in Sec. \ref{sec;Illustrative models}.
The vacuum we have obtained at the one-loop level is located at the point with non-vanishing $X_0$ and $X_{\pm}$. 
There, the D-term is suppressed by $|z|^2$ and 
the one-loop correction given by the non-vanishing F-terms can easily stabilize the vacuum.
The distance between the two SUSY breaking vacua is enough large for our vacuum to be long-lived, because of the runaway behavior. Thus, we focus on this vacuum and construct some models with the GUT gauge symmetries.

%Note that we can realize a large hierarchy between the U(1) gauge symmetry breaking scale and 
%the gaugino mass, depending on the ratio among the parameters: $f^2/m^2_1 m_2^2$, $\lambda$ and the gauge couplings. In fact, we will see that the condition, $\lambda f \gg m_1m_2$, that is consistent with the SUSY breaking vacuum, can realize a large hierarchy between the gaugino mass and the U(1) breaking scale,
%in the next section. 
Before the application to the GUT models, let us comment on the theoretical aspects of our SUSY breaking model.
Above, we have shown that the runaway direction can be lifted up 
by one-loop effects in one concrete model.
The runway behavior is the generic feature in a certain class of 
SUSY breaking models with gauge symmetries as explained in the previous section.
Similarly, runaway directions in generic models could be stabilized by loop-effects in 
proper parameter regions. 
It would be important to discuss conditions on lifting of runaway directions in generic models, 
but it is beyond our scope.

Here, we also give a comment on the R-symmetry.
The above model has the R-symmetry, whose charges are assigned such that 
$X_0$ and $X_\pm$ have the R-charge 2 and $\phi_\pm$ have vanishing charge.
At the minimum studied above, the fields $X_0$ and $X_\pm$ develop VEVs, and then the R-symmetry is spontaneously broken.
Note that the U(1) charges of $X_0$ and $X_\pm$ are different.
For example, if a VEV of a single field breaks the R-symmetry and U(1) symmetry, 
a new R-symmetry, which is a linear combination of the R-symmetry and 
broken U(1) symmetry, would remain.
However, in the above model, such a new R-symmetry does not remain.
Then, the gauge messenger contribution produces non-vanishing gaugino masses at the one-loop level.
We see the predictions in some illustrative models.

So far, we have studied the SUSY breaking model with the U(1) gauge symmetry.
We can extend this model to the model with non-Abelian gauge symmetry $G$.
In the next section, we apply the above study to models with non-Abelian gauge symmetry,
and discuss the applications to the Pati-Salam Model and the SU(5) GUT.

\section{Non-Abelian gauge models}
\label{sec3}

In this section, we extend the previous discussion on U(1) models to 
non-Abelian gauge models.

\subsection{$SU(N)$ model}

Here, we consider the extension of the U(1) model to non-Abelian gauge theory.
We replace $X_+$ and $\phi_+$ by chiral matter fields with $R$ representation under 
non-Abelian gauge symmetry ($G$), and 
$X_-$ and $\phi_-$ by chiral matter fields with conjugate representation, $\overline R$, 
while $X_0$ is the singlet.
For concreteness, we study the model with $G=SU(N)$ gauge symmetry, where 
$X_+$ and $\phi_+$ are the $N$ fundamental representations and 
$X_-$ and $\phi_-$ are its conjugate representations.
Then, we consider the same superpotential as Eq.(\ref{eq:W-model}) with 
the above replacement of representations.
Similar to Eqs.~(\ref{eq:pot-min}) and (\ref{eq:WX}), some components 
in $N + \bar N$ representations develop VEVs and non-vanishing F-terms.
By using $SU(N)$ rotation, we can fix the VEV directions as 
\begin{equation}
\phi^{(0)}_+ = \left(
\begin{array}{c}
0 \\
\vdots \\
0\\
\hat \phi^{(0)}_+
\end{array}
\right), \qquad 
\phi^{(0)}_- = \left(
\begin{array}{c}
0 \\
\vdots \\
0\\
\hat \phi^{(0)}_-
\end{array}
\right),
\end{equation}
with 
  \begin{equation}
\label{eq:pot-min-hat}
\hat \phi^{(0)}_+ = -\frac{F}{m_1}, \qquad \hat \phi^{(0)}_- = \frac{F}{m_2} .
\end{equation}
Thus, the gauge symmetry $SU(N)$ is broken to $SU(N-1)$.
Similarly, we obtain non-vanishing F-terms along the following directions: 
\begin{equation}
W_{X_+} = \left(
\begin{array}{c}
0 \\
\vdots \\
0\\
 W_{\hat X_+}
\end{array}
\right), \qquad 
W_{X_-} = \left(
\begin{array}{c}
0 \\
\vdots \\
0\\
 W_{\hat X_-}
\end{array}
\right),
\end{equation}
with 
\begin{equation}
\label{eq:WX-hat}
 W_{\hat X_+}=- W_{\hat X_-}=F.
\end{equation}
The F-term, $W_{X_0}$, is the same as Eq.(\ref{eq:WX}).

The D-terms corresponding to the broken generators are non-vanishing 
at $X_0=X_\pm =0$, but the tree-level potential has a runaway direction, which is the same as Eq.(\ref{eq:flat-ex-2}). 
Furthermore, similar to Eqs.(\ref{eq:V-1loop}), (\ref{eq:V-full}), and (\ref{eq:V-full-2}), the potential 
including one-loop effects would be written as
\begin{equation}\label{eq:V-full-21}
V = \frac{C_{min}}{|z|^2} + \frac{m^2_1m_2^2 \gamma_{X_0}}{\lambda^2} \ln \left (  |z|^2 F_0^2/\Lambda^2 \right )+\cdots .
\end{equation}
Here, $\gamma_{X_0}$ denotes the anomalous dimension of $X_0$, which depends on the coupling and 
multiplicity $N$.
Then, the minimum is estimated as 
\begin{equation}
|z^{(0)}|^2 = \frac{\lambda^2 C_{min}}{m_1^2m_2^2 \gamma_{X_0}}.
\end{equation}
Note that the gauge symmetry breaking scale is given by
\begin{equation}
M_X \sim g X^{(0)}_{\pm}.
\end{equation}

Compared to the U(1) model, there are extra fields from the decomposition of $\phi_\pm$ and $X_\pm$.
The VEVs of $X_0$, $X_{\pm}$ and $\phi_{\pm} $ can make the remnant fields massive at the tree level, except for $z$.
Then, we obtain the SU(N-1) gauge theory, effectively. 
Integrating out the remnant fields at the breaking scale, the mass of the SU(N-1) gaugino 
is radiatively induced. In addition, the mass squared of extra fields charged under SU(N)
would be also generated. In order to check the stability of our vacuum,
we need estimate the soft SUSY breaking terms. Below, we study the stabilities in some concrete models.

%Hence, we can realize the gauge symmetry breaking and SUSY breaking at the same time.
%Moreover, there is a hierarchy between their breaking scales, i.e., $M_X$ and $W_{X_{\pm}}/X_{\pm}$.
%Their ratio depends on the parameters.

Similarly, 
we can construct a model, where $SU(N)\times U(1)$ is broken by fields with $N_q$ representation 
and its conjugate where $q$ is U(1) charge.
Also we can construct a model, where $SU(N) \times SU(M)$ gauge symmetries are broken
by the fields with $(N, \, \overline M)$ representation and its conjugate, respectively.
Such models would be interesting from the viewpoint of phenomenological applications:
the flipped $SU(5)\times U(1)$ model \cite{flippedsu5} and $SU(4) \times SU(2)_L \times SU(2)_R$ model \cite{PS} could
correspond to the case.
In the next section, we discuss the application including quark and lepton chiral superfields
and study the soft SUSY breaking terms in each model.
Inclusion of squarks and sleptons, however, makes the potential complicated 
and in general there are directions, where squarks and sleptons develop their VEVs.
If all of the squark and slepton masses squared are positive, such a vacuum would be 
(meta-)stable.
We assume that quarks and leptons have no couplings with 
$X_0$, $X_\pm$ and $\phi_{\pm}$.
Then, we estimate soft SUSY breaking terms through the gauge mediation.
We give some comments on the stability of our vacuum in each setup.

\subsection{Illustrative models}
\label{sec;Illustrative models}

Based on the above discussion, we construct illustrative models where 
gauge symmetry and SUSY are simultaneously broken.
In the previous section, we introduce the extension to the model with SU(N) gauge symmetry.
In the same manner, we can consider a model with $G_1 \times G_2$ gauge symmetries as well.
Here, $G_{A}$ ($A=1,\,2$) is Abelian or non-Abelian gauge symmetry, and
both $X_{\pm}$ and $\phi_{\pm}$ are charged under $G_1 \times G_2$, while
$X_0$ is the singlet.

In our SUSY breaking model, the VEVs of $X_{\pm}$ and $\phi_{\pm}$ break gauge symmetry.
If $G_1 \times G_2$ has a bigger rank than the SM gauge symmetry,
we could discuss the simple scenario that the SM gauge symmetry is embedded into $G_1$ and/or $G_2$
like the GUT and the SUSY breaking sector also causes the GUT breaking.
Since the dynamics of SUSY breaking and GUT breaking is explicitly given in this kind of model, 
the soft SUSY breaking terms for the supersymmetric SM fields are explicitly predicted according to the gauge mediation.
Thus, in this subsection, we evaluate soft SUSY breaking terms from the gauge mediation.
We neglect D-term contributions in the study below.

Let us assume that one of the SM gauge groups ($G^{{\rm SM}}_a$) is given by the part of $G_1\times G_2$, the gaugino mass of $G^{{\rm SM}}_a$ is generated at the gauge symmetry breaking scale $\mu$ as \cite{Giudice:1997ni}
\begin{equation}
\label{eq;gaugino}
M_a (\mu) =  \frac{\alpha_a (\mu)}{4 \pi} \Delta b_a \frac{F^X}{X}.
\end{equation}
Here, $\Delta b_a$  denotes the difference between the beta-function coefficients of $G^{{\rm SM}}_a$ and of $G_1 \times G_2$.\footnote{In our notation, $G^{{\rm SM}}_a$ ($a=1,\,2,\,3$) represents $G^{{\rm SM}}_1\equiv U(1)$, $G^{{\rm SM}}_2\equiv SU(2)_L$ and $G^{{\rm SM}}_3 \equiv SU(3)$, respectively. Each of the beta-function coefficient in the MSSM is
denoted by $b_1$, $b_2$ and $b_3$. Note that the $U(1)$ gauge coupling is the one of the unified gauge couplings around $10^{16}$ GeV in the MSSM. The beta-function of $U(1)_Y$ is denoted by $b_Y$. }
For instance,
if the SM $SU(3)$ comes from $G_1$, $\Delta b_3$ is given by
$\Delta b_3=b_3-b'_1,$ where $b_3$ and $b'_1$ are the beta-function coefficients of $SU(3)$ and $G_1$, respectively. 
Here, we assume that chiral superfields integrated out at $\mu$ obtain the masses from
the non-vanishing VEV, $X$. $F^X$ is the F-term of the superfield developing the VEV.

When the MSSM chiral superfield, $Q_I$, is charged under $G_1\times G_2$,
the non-vanishing A-term and B-term are generated as follows \cite{Giudice:1997ni}:
\begin{eqnarray} \label{eq;A-term}
A_{I}  (\mu)&=&  \frac{1}{2 \pi} \left \{c^A_I  \alpha_A(\mu)  -c^a_I  \alpha_a(\mu)  \right \} \frac{F^X}{X}, 
\end{eqnarray}
Here, $c^A_I$ and $c^a_I$ are the second Casimir operators of $G_A$ and $G^{{\rm SM}}_a$.
The SUSY breaking trilinear coupling corresponding to the Yukawa coupling $y_{IJK}$, $y_{IJK}A_{IJK}Q_IQ_JQ_K$, and the SUSY breaking bilinear coupling corresponding to the $\mu$-term, $\mu_H B H_u H_d$, are given by $A_{IJK}=A_I+A_J+A_K$ and $B=A_{H_u}+A_{H_d}$. Note that $H_u$ and $H_d$ denote the $SU(2)_L$-doublet Higgs fields in the MSSM.

It is a critical feature of this model that non-vanishing A-terms and B-term are generated at the one-loop level.
In order to realize 125 GeV Higgs mass, a sizable A-term involving top squark is
favorable. Besides, a proper value of the B-term is also necessary to cause the EW symmetry breaking.
Then, this feature would be appropriate to construct a realistic supersymmetric model.

%On the other hand, we may suffer from tachyonic masses.
Next, we estimate the scalar mass squared in our model.
As discussed in Ref. \cite{Intriligator:2010be}, there are one-loop corrections to the scalar masses squared in this kind of supersymmetric model. In our model, $\phi_{\pm}$ and their F-terms also develop non-vanishing VEVs, and
the VEVs drive the masses squared negative according to the one-loop level \cite{Intriligator:2010be}.
We estimate the one-loop corrections as
\begin{equation}
m_I^2  (\mu) =  - \frac{1}{2 \pi}  \left \{c^A_I  \alpha_A(\mu)  -c^a_I \alpha_a(\mu) \right \} 
 {\cal M}^2_{1} ,
\end{equation}
where ${\cal M}^2_{1}$ is given by
\begin{equation}
{\cal M}^2_{1}= \frac{\left ( |\Hat \phi^{(0)}_{+}|^2+|\Hat \phi^{(0)}_{-}|^2+ 2 |z^{(0)}|^2F^2 \right ) \left( |F_{\Hat \phi_{+}}|^2+|F_{\Hat \phi_{-}}|^2+2 F^2 \right ) - \left | \Hat \phi^{(0)}_{+}F_{\Hat \phi_{+}}+ \Hat \phi^{(0)}_{-}F_{\Hat \phi_{-}} +2 z^{(0)} F^2   \right |^2}{ \left ( |\Hat \phi^{(0)}_{+}|^2+|\Hat \phi^{(0)}_{-}|^2+ 2 |z^{(0)}|^2F^2 \right )^2}.
\end{equation}
%assuming $|\phi_{\pm}|^2 \ll |X_{\pm}|^2$ to realize our SUSY breaking vacuum.
%\footnote{
%If this condition is not satisfied, the estimations in Eqs. (\ref{eq;app1}) and (\ref{eq;app2})
%are spoiled and especially the directions of $\phi_\pm$ become unstable.}
${\cal M}^2_{1}$ is vanishing in the limit that $|\Hat \phi^{(0)}_{\pm}|$ and $F_{\Hat \phi_{\pm}}$ go to zero.
In our model, $|\Hat \phi^{(0)}_{\pm}|$ is relatively small compared to $|\Hat X^{(0)}_{\pm}|$. $F_{\Hat \phi_{\pm}}$ is also suppressed by $|z^{(0)}|$ as shown in Eq. (\ref{eq;evaluation}), 
%We can actually see this condition at the reference point in Eq. (\ref{eq;evaluation}).
so that the one-loop corrections are expected to be small.
For instance, ${\cal M}^2_{1}$ is estimated as $(2.5 \times 10^{-5})/|z^{(0)}|^2$,
at the reference point in Eq. (\ref{eq;evaluation}). 
On the other hand, the two-loop contributions to the masses squared are estimated as
\begin{eqnarray} 
%A_{I}  (\mu)&=&  \frac{1}{2 \pi} \left \{c^A_I  \alpha_A(\mu)  -c^a_I  \alpha_a(\mu)  \right \} \frac{F^X}{X}, \\
m_I^2  (\mu) &=&   \frac{1}{8 \pi^2}  \left \{c^A_I b'_A \alpha^2_A(\mu)  +c^a_I \widetilde b_a  \alpha^2_a(\mu) \right \} 
 \left( \frac{F^X}{X} \right)^2. \label{eq;squaredmass}
\end{eqnarray}
Note that $\widetilde b_a$ is given by $\widetilde b_a= b_a-2b'_1$, when $G^{{\rm SM}}_a$
is a subgroup of $G_1$.
Here, $F^X/X$ is dominantly given by $X^{(0)}_{0}$ and $X^{(0)}_{\pm}$ and estimated as $1/|z^{(0)}|$.
Thus, the two-loop contributions could dominate over the one-loop, as far as the gauge couplings are not too small.  
At the reference point in Eq. (\ref{eq;evaluation}), the minimum size of the gauge coupling is about $0.04$ for the two-loop correction to be dominant compared to the one-loop.
Note that the one-loop contribution is suppressed more significantly, if $F$ is assumed to be much larger than $m^2_{1,2}$.

%%%%%%%%%%%%%%%%%%%%%%%%%

Even if the two-loop contributions dominate the masses squared, the beta-function coefficient of $G_A$ 
may give a negative contribution to the masses squared, as shown in Eq. (\ref{eq;squaredmass}).
In such a case, we would conclude that the vacuum is not stable, when only the gauge mediation is dominant.
We need additional contributions to sfermion masses, e.g. gravity mediation, unless large RG corrections are expected.
We will give a comment on the extra contributions in Sec. \ref{sec;tachyon}.

Below, we especially introduce two different models: the Pati-Salam model \cite{PS} and the $SU(5) \times U(1)$ GUT, namely the flipped $SU(5)$ GUT \cite{flippedsu5}. In each model, we show the soft SUSY breaking terms and 
discuss the phenomenological impacts. We also give a short discussion about the conventional $SU(5)$ model \cite{su5}.
Concerned with the soft SUSY breaking terms, we investigate the one-loop corrections for the gaugino and the A-terms
and especially the two-loop corrections for the mass squared. The one-loop corrections may be dominant, depending on the parameters. The one-loop, however, gives negative mass squared, so that we discuss the possibility that
the two-loop corrections to the mass squared compensate the tachyonic mass in each model.

\subsubsection{Pati-Salam model}

First, we apply our SUSY breaking dynamics to the Pati-Salam model with the gauge symmetry $SU(4) \times SU(2)_R \times SU(2)_L$ \cite{PS}. In the Pati-Salam model, $SU(4) \times SU(2)_R$ breaks down to $SU(3) \times U(1)_Y$:
$SU(3)$ comes from the subgroup of $SU(4)$, and $U(1)_Y$ is given by the 
linear combination of the subgroups of $SU(4)$ and $SU(2)_R$.
In this case, $SU(4) \times SU(2)_R$ corresponds to $G_1 \times G_2$ in the above discussion.
The charge assignment of $SU(4) \times SU(2)_R \times SU(2)_L$ for $X_\pm$ and $\phi_\pm$
is defined as
\begin{equation}
\label{eq;PS}
X_+, \, \phi_+:(\bf{4,\,2,\,1}),~~~~X_-, \, \phi_-:(\overline{\bf{4}},\,\bf{2},\,\bf{1}).
\end{equation} 
$X_0$ is not charged under any gauge symmetry.
In addition to these, we set three generations of the usual Pati-Salam model, that 
correspond to $({\bf 4,\,1,\,2})$ and $({\bf \bar4,\,2,\,1})$ under $SU(4) \times SU(2)_R \times SU(2)_L$ 
as well as the Higgs fields corresponding to $({\bf 1,2,2})$.

Based on the study in Sec. \ref{sec;model}, we can expect that 
the VEVs of $X_{\pm}$ and $\phi_\pm$ break $SU(4) \times SU(2)_R$ at the SUSY breaking vacuum.
The remnant symmetry is expected to be $SU(3) \times U(1)_Y$ in the setup, so 
that our SUSY breaking model in Sec. \ref{sec;model} is compatible with the Pati-Salam model.

In our model, all fields from $X_\pm$ and $\phi_\pm$ can
gain the masses around the SUSY breaking scale.
Then, $\Delta b_{a}$ are evaluated as follows, assuming that the chiral superfields in the SUSY breaking sector
are integrated out at $\mu$:
\begin{equation}
(\Delta b_{Y}, \, \Delta b_{2}, \, \Delta b_{3})=(-10/3, \,0,\,-1).
\end{equation}
These values lead vanishing wino mass and relatively small gluino mass, according to 
Eq. (\ref{eq;gaugino}).

Following Eq. (\ref{eq;A-term}) and Eq. (\ref{eq;squaredmass}), the A-terms and masses squared
are also evaluated. We see that non-vanishing A-terms are generated,
if $Q_I$ is charged under $SU(4) \times SU(2)_R$.
In the mass squared, the signs of $b'_A$ and $b_a$ play a crucial role
in avoiding the tachyonic masses. In our setup, 
$X_\pm$ and $\phi_\pm$ largely contribute to the beta-function coefficients of $SU(4) \times SU(2)_R$:
$b'_{SU(4)}=4$ and $b'_{SU(2)_R}=-8$. Also, $\widetilde b_3=-5$ is led by
this matter content, so that the masses squared of right-handed squarks tend to be
negative.
The soft-SUSY breaking terms relevant to down-type and up-type squarks are obtained as follows:
\begin{eqnarray}
A_{Q_L u_R H_u}(\mu)&=& \left ( \frac{13}{6} \frac{\alpha_3(\mu)}{4 \pi} -\frac{13}{9} \frac{\alpha_Y(\mu)}{4 \pi} + 3 \frac{\alpha_R(\mu)}{4 \pi} \right ) \frac{F^X}{X}, \\
m^2_{Q_L}(\mu)&=& \left \{  \frac{5}{3} \frac{\alpha^2_3(\mu)}{(4 \pi)^2}+\frac{13}{54} \frac{\alpha^2_Y(\mu)}{(4 \pi)^2}  \right \} \frac{|F^X|^2}{|X|^2}, \\
m^2_{u_R}(\mu)&=&\left \{  \frac{5}{3} \frac{\alpha^2_3(\mu)}{(4 \pi)^2}+\frac{104}{27} \frac{\alpha^2_Y(\mu)}{(4 \pi)^2} -12 \frac{\alpha^2_R(\mu)}{(4 \pi)^2}  \right \} \frac{|F^X|^2}{|X|^2}, \\
m^2_{d_R}(\mu)&=& \left \{  \frac{5}{3} \frac{\alpha^2_3(\mu)}{(4 \pi)^2}+\frac{26}{27} \frac{\alpha^2_Y(\mu)}{(4 \pi)^2} -12 \frac{\alpha^2_R(\mu)}{(4 \pi)^2}  \right \} \frac{|F^X|^2}{|X|^2}.
\end{eqnarray}
Here, $Q_L$, $u_R$, and $d_R$ denote the $SU(2)_L$-doublet, $SU(2)_L$-singlet up-type, and down-type quark superfields respectively.
In these descriptions, the gauge coupling of $SU(4)$ is the same as the one of the SM $SU(3)$.
In addition, $\alpha_{R}(\mu)$ denotes the gauge coupling of $SU(2)_R$ symmetry, and
satisfies the following relation at the breaking scale;
\begin{equation}
\label{eq;gaugecouplingR}
\alpha^{-1}_Y(\mu)=\alpha^{-1}_R(\mu)+ \frac{2}{3} \, \alpha^{-1}_3(\mu),
\end{equation}
where $\alpha_Y$ denotes the $U(1)_Y$ gauge coupling.
As we see, the sizable $\alpha_R(\mu)$ gives the negative contributions to
$m^2_{d_R}$ and $m^2_{u_R}$. 
Depending on the breaking scale,
$\alpha_R$ becomes compatible with $\alpha_3$ and makes $m^2_{d_R}$ and $m^2_{u_R}$ negative.
This means that up-type and down-type squarks become tachyonic at the low scale even if the two-loop contributions are dominant,
as far as large positive RG corrections are not expected.
In this model, the gluino mass is relatively light, so that 
the RG correction is relatively small.

In the mass squared for right-handed slepton, there is also a negative contribution from $SU(2)_R$:
\begin{equation}
m^2_{e_R} (\mu)= \left \{ \frac{26}{3} \frac{ \alpha_1^2 (\mu)}{(4\pi)^2} + 15  \frac{ \alpha_3^2 (\mu)}{(4\pi)^2} - 12  \frac{ \alpha_R^2 (\mu)}{(4\pi)^2} \right \} \frac{|F^X|^2}{|X|^2}.
\end{equation}
The $SU(4)$ gauge interaction, however, compensates for the negative contribution,
so that $m^2_{e_R}$ can become larger than $m^2_{d_R}$ and $m^2_{u_R}$.
Note that the mass squared for left-handed lepton is also positive, because of no $SU(2)_R$ contribution.

We conclude that this application of our SUSY breaking scenario to the Pati-Salam model
works well to cause both SUSY breaking and GUT breaking.
The R-symmetry is spontaneously broken, so that finite gaugino masses are generated by the gauge mediation.
This model may, however, suffer from the tachyonic squark masses, if the gauge mediation contribution is dominant
in the soft SUSY breaking terms.
If the breaking scale is lower than $10^{10}$ GeV, all masses squared can be positive 
because of small $\alpha_R$.
Otherwise, we need other sizable mediation effects such as gravity mediation and anomaly mediation,
to lead a realistic supersymmetric SM model. The vanishing wino mass also requires such effects.

\subsubsection{flipped SU(5) GUT}

Next, we consider another application of our SUSY breaking scenario to the GUT model:
$G_1 \times G_2 \equiv SU(5)\times U(1)_X$. 
If $U(1)_Y$ is given by the linear combination of $U(1)_X$ and the subgroup of $SU(5)$,
the GUT model could correspond to the flipped SU(5) GUT \cite{flippedsu5}.
In the flipped $SU(5)$ GUT, we consider the charge assignment of $SU(5)\times U(1)_X$ for $X_\pm$ and $\phi_\pm$ as follows:
\begin{equation}
\label{eq;SU(5)}
X_+, \, \phi_+:({\bf{10}}, \, 1/\sqrt{40}),~~~~X_-, \, \phi_-:(\overline{\bf{10}},\, -1/\sqrt{40}).
\end{equation} 
$X_0$ is again not charged under any gauge symmetry.
In this GUT, the MSSM fields are again embedded into 
${\bf 10}$, ${\bf 5}$ and $\overline{{\bf 5}}$ representational fields, and the GUT breaking should consist of
$SU(5) \to SU(3) \times SU(2) \times U(1)_5$ and $U(1)_X\times U(1)_5 \to U(1)_Y$.
The SUSY breaking vacuum discussed in Sec. \ref{sec;model} leads the breaking chain.
Note that ${\bf 5}$ and $\overline{{\bf 5}}$ representational fields, denoted by $H$ and $\overline H$ respectively, 
are also introduced
to realize the EW Higgs doublets in the MSSM. In order to avoid too short life time of proton,
the masses of the colored fields in the ${\bf 5}$ and $\overline{{\bf 5}}$ Higgs fields should be
GUT-scale. In our setup, we can write down the following terms: $X_+ X_+ H$, $X_+ \phi_+ H$, $X_- X_- \overline{H}$,
$X_- \phi_- \overline H$, and so on.\footnote{These terms do not modify our vacuum, since
the SM-singlet fields in $X_\pm$ and $\phi_\pm$ only develop the VEVs and the linear terms such as $\langle X_- \rangle \langle X_+ \rangle H$ are vanishing at our vacuum.}
Then, we expect that the colored Higgs fields 
can obtain the masses around the GUT scale, and mediate the SUSY breaking to the visible sector.
Note that we may have to assign R-symmetry to the visible sector and we need 
some mechanisms to generate the low-scale $\mu$ term, that is the supersymmetric mass term
of the Higgs doublets. This issue is beyond our scope, and
we estimate the soft SUSY breaking terms assuming that the colored Higgs fields are also
integrated out at the GUT breaking scale and mediate the SUSY breaking effect.

The threshold corrections, that correspond to the coefficients of the gaugino masses,
are given by  
\begin{equation}
(\Delta b_{1}, \, \Delta b_{2}, \, \Delta b_{3})=(1, \,-3,\,1).
\end{equation}
Note that $\Delta b_{3}$ is relatively small, and it is vanishing if the colored Higgs fields 
do not contribute to the soft SUSY breaking term.

In this GUT model, the beta-function coefficients are not so large:
$b'_{SU(5)}=2$ and $b'_{U(1)_X}=-8$.  The coefficient, $\widetilde b_3$, that appear in the mass squared for squark, 
is estimated as $\widetilde b_3=-1$. 
The soft SUSY breaking terms concerned with the squark and slepton masses are estimated as follows:
\begin{eqnarray}
A_{Q_L u_R H_u}(\mu)&=& \left( \frac{127}{15} \, \frac{\alpha_3(\mu)}{4 \pi}+\frac{7}{10} \, \frac{\alpha_X(\mu)}{4 \pi}-\frac{13}{15} \, \frac{\alpha_1(\mu)}{4 \pi} \right ) \frac{F^X}{X}, \\
m^2_{Q_L}(\mu)&=&\left \{ \frac{127}{30} \, \frac{\alpha^2_3(\mu)}{(4 \pi)^2}-\frac{2}{5} \, \frac{\alpha^2_X(\mu)}{(4 \pi)^2}+\frac{43}{150} \, \frac{\alpha^2_1(\mu)}{(4 \pi)^2} \right \} \frac{|F^X|^2}{|X|^2},  \\
m^2_{u_R}(\mu)&=&\left \{\frac{104}{15} \, \frac{\alpha^2_3(\mu)}{(4 \pi)^2}-\frac{18}{5} \, \frac{\alpha^2_X(\mu)}{(4 \pi)^2}+\frac{344}{75} \, \frac{\alpha^2_1(\mu)}{(4 \pi)^2} \right \}  \frac{|F^X|^2}{|X|^2}, \\
m^2_{d_R}(\mu)&=&\left \{\frac{176}{15} \, \frac{\alpha^2_3(\mu)}{(4 \pi)^2}-\frac{2}{5} \, \frac{\alpha^2_X(\mu)}{(4 \pi)^2}+\frac{86}{75} \, \frac{\alpha^2_1(\mu)}{(4 \pi)^2} \right \} \frac{|F^X|^2}{|X|^2}, \\
m^2_{L}(\mu)&=&\left \{ \frac{21}{10} \, \frac{\alpha^2_3(\mu)}{(4 \pi)^2}-\frac{18}{5} \, \frac{\alpha^2_X(\mu)}{(4 \pi)^2}+\frac{129}{50} \, \frac{\alpha^2_1(\mu)}{(4 \pi)^2} \right \} \frac{|F^X|^2}{|X|^2}, \\
m^2_{e_R}(\mu)&=&\left \{ -10 \, \frac{\alpha^2_X(\mu)}{(4 \pi)^2}+\frac{258}{25} \, \frac{\alpha^2_1(\mu)}{(4 \pi)^2} \right \} \frac{|F^X|^2}{|X|^2}. \label{meR-flipped}
\end{eqnarray}
Here, $\alpha_1$ and $\alpha_X$ satisfy the following relation,
\begin{equation}
\label{eq;gaugecouplingX}
25 \, \alpha^{-1}_1(\mu)=24 \, \alpha^{-1}_X(\mu)+   \alpha^{-1}_3(\mu).
\end{equation}
Note that $m^2_{L}$ is the mass squared for left-handed slepton.
When $\mu$ is set to the GUT scale ($\sim 10^{16}$ GeV), all gauge couplings get
close to the same value. If the couplings are assumed to be unified at $\mu$,
we find that the two-loop contributions to all masses squared of squarks and sleptons can be positive at the breaking scale in this GUT model. Note that the gauge couplings are also enough large to compensate the negative contributions of the one-loop to the masses squared.

In our analysis, we have not included the threshold correction that arises from the mass difference of 
the particles in SUSY breaking sectors. Besides, we have not detailed the setup for the realistic model.
For instance, we have to take into account how to realize the Yukawa couplings in the MSSM.
If we introduce extra fields to build a realistic model, the predictions we obtained here
would be modified. The detailed analysis will be given near future.

Let us comment on the not-flipped SU(5) GUT case \cite{su5}. 
In this case, the gauge symmetry consists of two symmetry: $G_1 \times G_2 \equiv SU(5)\times U(1)^\prime$.
$U(1)_Y$ comes from the subgroup of $SU(5)$ and we could, for instance, consider the following charge assignment for the SUSY breaking sector:
\begin{equation}
\label{eq;SU(5)}
X_+, \, \phi_+:({\bf{adj}}, \, 1),~~~~X_-, \, \phi_-:({\bf{adj}},\, -1).
\end{equation} 
This setup, however, leads very large negative $b'_{SU(5)}$, because of many adjoint chiral superfields:
$b'_{SU(5)}=-12$. This large value leads Landau pole just above the breaking scale.
Besides, we face the big issue concerned with the masses of the colored Higgs fields.
In the $SU(5)$ GUT, we introduce two terms, $W_H=\mu_H H \overline H+\lambda_\Sigma \Sigma H \overline H$,
where $ \Sigma $ is the adjoint field to break the $SU(5)$ gauge symmetry.
We have to allow the fine-tuning between $\mu_H$ and $\lambda_\Sigma \langle \Sigma \rangle $, but
in principal we obtain the large hierarchy between the EW Higgs doublet and the colored Higgs fields.
Now, we can expect that either $X_\pm$ or $\phi_\pm$ plays a role of $\Sigma$
and realizes the hierarchy.  The $U(1)^\prime$ symmetry, however, forbids either
$\mu_H$ or $\lambda_\Sigma$, so that it is impossible to gain the hierarchy in this setup.
Besides, the VEVs of $X_\pm$ are expected to be large, and then 
$X_\pm$ should be identical to $\Sigma$ in $W_H$.
This setup, however, causes the bilinear term of the scalar components of $H$ and $\overline H$,
according to the non-vanishing F-terms of $X_\pm$.
Therefore, it is difficult to realize the realistic EW symmetry breaking vacuum.

\subsection{Tachyonic mass}
\label{sec;tachyon}

We have studied two examples towards constructing realistic models.
Indeed, by the mechanism in section \ref{sec:SUSY-breaking}, we 
can break the gauge symmetry and SUSY in realistic GUT gauge theories.
However, only pure gauge mediation may lead to tachyonic squark and/or slepton masses, especially in the Pati-Salam model. 
That implies that such vacua are not stable or even meta-stable.
In order to stabilize the vacuum, we need another contribution, e.g. gravity mediation, in such a case.
For example, we can assume the additional term in K\"ahler potential,
\begin{equation}
\Delta K = (c_0|X_0| + c_{+}|X_+|^2 + c_-|X_-|^2) |Q_I|^2,
\end{equation}
where $Q_I$ denotes quark and lepton superfields,
such that squarks and sleptons have positive masses squared.
Phenomenological aspects of models depend strongly on $c_0$ and $c_\pm$.
On the other hand, if we have no additional corrections on 
the gaugino masses and $A$-terms except the pure gauge mediation, 
these can be predictions of our models.
Alternatively, we may assume that the anomaly mediation 
\cite{Randall:1998uk}
is comparable with 
the gauge mediation discussed above.
The pure anomaly mediation leads to tachyonic slepton masses, although squark masses squared are 
positive. 

In the Pati-Salam model, the vanishing wino mass also requires such additional contributions.
A proper combination of the gauge mediation, the gravity mediation and anomaly mediation would lead 
realistic mass spectrum of the SUSY particles in a certain GUT breaking model.
Such a study is challenging and we would study it elsewhere.

\section{Conclusion}   \label{conclusion}

It is one of important issues to understand the vacuum structure of our universe.
If SUSY really exists in our nature, our vacuum spontaneously breaks the symmetry, 
so that it is a major issue to construct a SUSY breaking model.

When SUSY breaking is triggered by F-terms of chiral superfields,
it is known that the symmetry breaking is accompanied by flat directions in the field space.
The flat directions should be stabilized at the non-vanishing VEV to realize the R-symmetry breaking.
Besides, it is a big issue to induce non-vanishing gaugino masses in the gauge-mediation models, 
even if the R-symmetry is broken at our vacuum. Thus, it is not trivial to find the realistic SUSY breaking vacua and construct the SUSY model that predicts massive superpartners of the SM particles.

In this paper, we consider a supersymmetric model with U(1) gauge symmetry and R-symmetry.
In this model, both of the gauge symmetry and SUSY are broken by the same fields. 
We find flat directions triggered by the SUSY breaking, and the D-term of the U(1) gauge symmetry 
is not vanishing along the flat directions. In this kind of model, it is known that there are also runaway directions at the tree-level \cite{Azeyanagi:2011uc}.
We suggest that such runaway directions can be lifted by the one-loop effect, and the SUSY breaking vacuum can be realized. The gauge symmetry breaking is also caused by the SUSY breaking dynamics,
and the R-symmetry also spontaneously breaks down.
In such a case, the gauge messenger field can mediate the SUSY breaking effect and can induce non-vanishing gaugino masses.

We can extend this U(1) model to non-Abelian theory.
It is quite interesting to apply this mechanism to the GUTs, e.g. the Pati-Salam model and the flipped SU(5).
This simple setup may, however, cause the problem that squarks and sleptons develop VEVs according to the one-loop and two-loop corrections.
We estimate the soft SUSY breaking terms concerned with sfermions through the gauge mediation.
In the Pati-Salam model, the $SU(2)_R$ contributions to the mass squared are negative even at the two-loop level, so that
especially the squark masses become tachyonic depending on the size of gauge coupling, i.e. the breaking scale. 
On the other hand, we find that all masses squared can be positive in the flipped $SU(5)$, taking into account the two-loop corrections.
We need study in more detail, taking into account how to realize the realistic Yukawa couplings in the MSSM.
In the case that the negative mass squared is derived, we propose another contribution, e.g. gravity mediation and anomaly mediation. 
In particular, such additional contributions are required by the vanishing wino mass in the Pati-Salam model.
Those contributions may drastically change the mass spectrum, and phenomenology may depend on details of mediations.
Further study on the GUT with our SUSY breaking model will be given in the near future.

%======================================%
%<<<<<<<<<< ACKNOWLEDGMENTS >>>>>>>>>>>%
%======================================%

\section*{Acknowledgments}

This work is supported in part by the Grant-in-Aid for Scientific Research 
 No.~26247042 and No.~17H05395 (T.K.) and No. 17H05404 (Y.O.) from the Ministry of Education, Culture, Sports,
 Science and Technology in Japan. 
%

%======================================%
%<<<<<<<<<< APPENDIX >>>>>>>>>>>%
%======================================%
%\appendix%
%
%\section{} 

%======================================%
%<<<<<<<<<<< bibliography >>>>>>>>>>>>>%
%======================================%

%%%%%%%%%%%%%%%%%%%%%%%%%%%%%%%%%%%%%%%%%%%%%%%%%%%%%%%%%%%%

\end{document}